\documentclass[12pt]{article}
\usepackage{axodraw}
\usepackage{graphicx}

\newcommand{\bq}{\begin{equation}}
\newcommand{\eq}{\end{equation}}
\newcommand{\bqa}{\begin{eqnarray}}
\newcommand{\eqa}{\end{eqnarray}}
\newcommand{\nll}{\nonumber\\}

\newcommand{\als}{\alpha_{_S}}

\newcommand{\sss}[1]{\scriptscriptstyle{#1}}
\newcommand{\ds }{\displaystyle}
\def\mw {M_{\sss{W}}}
\def\mz {M_{\sss{Z}}}
\def\mup{m_{u}}
\def\mdn{m_{d}}

\def\ml{m_{\ell}}
\def\hs{\hat s}
\def\mwt {\widetilde{M}^2_{\sss{W}}}

\def\MSbar{$\overline{\mathrm{MS}}\ $}
\newcommand{\GeV}{\unskip\,\mathrm{GeV}}
\newcommand{\MeV}{\unskip\,\mathrm{MeV}}
\def\gw {\Gamma_{\sss W}}
\def\gz {\Gamma_{\sss Z}}
\def\mh {M_{\sss H}}
\def\order#1{{\mathcal O}\left(#1\right)}

\title{NLO QCD corrections to Drell-Yan processes in the {\tt SANC} framework}

\author{A.~Andonov$^1$,
A.~Arbuzov$^{2,3}$,
S.~Bondarenko$^{2,3}$, 
P.~Christova$^3$, V.~Kolesnikov$^3$, \\
G.~Nanava$^4$, R.~Sadykov$^3$ \\[.2cm]
{\it $^{1}$Bishop Konstantin Preslavsky University, Shoumen, Bulgaria }\\ 
{\it $^{2}$Bogoliubov Laboratory of  Theoretical Physics, JINR } \\ 
{\it $^{3}$Dzhelepov Laboratory of Nuclear Problems, JINR  }    \\
{\it        ul. Joliot-Curie 6, RU-141980 Dubna, Russia  }\\
{\it $^4$Physikalisches Institut der Universit\"at Bonn,} \\
{\it     Nussallee 12, D-53115,  Bonn, Germany}
}

\date{}

\begin{document}

\maketitle

\begin{abstract}
NLO QCD corrections to charged and neutral current Drell-Yan processes 
and their implementation in the computer system {\tt SANC} are considered. 
On the partonic level both quark-antiquark and quark-gluon 
scattering channels are taken  into account. Subtractions of the collinear 
singularities in the massive case are compared with ones in the 
\MSbar scheme.
Results of {\tt SANC} on the hadronic level are presented.
Comparison with results of the {\tt MCFM} package is shown.
\end{abstract}

\section{Introduction}
 
Charged and neutral current Drell--Yan (DY) processes~\cite{Drell:1970wh} on the eve of the 
first proton collisions at the LHC become very important for precision tests
of the Standard Model. They are easily detected and will provide standard 
candles for detector calibration during the first stage of LHC running.
They will be also used for extraction of partonic density functions (PDF)
in the kinematical region which has not been accessed by earlier experiments.
Therefore it is crucial to control the theoretical predictions for production
cross sections and kinematic distributions of these processes.   

In the previous paper~\cite{Andonov:2007zz} we  presented a part of
the QCD sector of our computer system {\tt SANC}~\cite{Andonov:2004hi} 
({\it http://sanc.jinr.ru/} and {\it http://pcphsanc.cern.ch/ })
where the NLO QCD processes are treated. 
There we considered the implementation into {\tt SANC} the calculation of 
the  charged (CC) and neutral (NC) current quark--antiquark
Drell--Yan processes on the partonic level and briefly presented
some numerical results for the hadronic level. 
The QCD corrections to DY processes are known in the literature
for many years, see Refs.~\cite{Altarelli:1979ub,KubarAndre:1978uy,Hamberg:1990np}. 
Recently the corresponding NNLO corrections for differential distributions
have been received~\cite{Anastasiou:2003ds,Melnikov:2006kv}.
 
In this paper with respect to Ref.~\cite{Andonov:2007zz}
we add into consideration quark--gluon and gluon-antiquark  
Drell--Yan processes on the partonic level side by side 
with the quark--antiquark ones. We implemented into {\tt SANC} the 
calculation of QCD corrections to DY processes on the hadronic level.
Working with massive quarks, we  regularized the collinear singularities  
by masses of quarks. But on the hadronic level we have to remove these 
collinear singularities to avoid double counting, because they are already 
included in PDF. Therefore we compared analytical results obtained in our 
treating of collinear singularities  with analogous results obtained in  
the \MSbar scheme  calculated in the $n$-dimensional phase space.
In this way we extracted the  subtraction terms needed to remove
the collinear singularities in our massive quarks case. 
We show here also comparison with the corresponding
results of the MCFM~\cite{Ellis:1999ec} package.

One-loop electroweak radiative corrections were computed for the
DY processes by {\tt SANC} in Refs.~\cite{Arbuzov:2005dd,Arbuzov:2007db,Arbuzov:2007kp} 
and extensively compared with results of other groups, 
see {\it e.g.} Refs.~\cite{Gerber:2007xk,Buttar:2008jx,Bardin:2008fn}.  

The paper is organized as follows.
In the second section we calculated the hard gluon bremsstrahlung 
contributions both to charged and neutral current Drell-Yan processes 
on a quark-parton level in the massless  \MSbar scheme 
and compared with our massive quarks results. We used 
{\tt FORM3.1}~\cite{Vermaseren} for analytical calculations.
In the third section we calculated the quark--gluon and gluon-antiquark
DY processes with charged and neutral currents in our massive 
quarks treating and draw parallel with calculation in the massless
\MSbar scheme. In Conclusion we discuss the results and give some illustrations
of DY distributions at the hadronic level.

\section{Quark--antiquark  Drell--Yan processes }

\subsection{Massive quarks treatment of the hard gluon contribution}

Working with massive quarks we showed in  the previous paper~\cite{Andonov:2007zz} 
the NLO QCD corrections due to the hard gluon bremsstrahlung 
of  charged current (CC) 
~$ {\bar d}(p_1)+ u(p_2) \to W \to {\nu}_\ell(p_3) + {\bar \ell}(p_4) + g(p_5) $
and  neutral current (NC) 
~${\bar q}(p_1)+ q(p_2) \to [A, Z] \to \ell(p_3) + {\bar \ell}(p_4) + g(p_5) $
Drell--Yan processes were obtained in the form
\bqa
  {\hat \sigma}^{\rm CC}_{\rm Hard} = \frac{\als}{2 \pi}
\, \int_{z_{min}}^{z_{max}}  d z \,
 \,{\hat \sigma}^{\rm CC}_0(z \hs) \,\, P_{qq}(z) \,\,
\left[2 \ln\left(\frac{\hs}{\mu^2}\right) +\ln\left(\frac{\mu^2}{\mup^2}\right)
      +\ln\left( \frac{\mu^2}{\mdn^2}\right)- 2\right], 
\label{HardcontribCC}
\eqa
\bqa
  {\hat \sigma}^{\rm NC}_{\rm Hard} = \frac{\als}{\pi}
\, \int_{z_{min}}^{z_{max}}  d z \,
{\hat \sigma}^{\rm NC}_0(z \hs) \,P_{qq}(z) \,\,
\left[ \ln\left(\frac{\hs}{\mu^2}\right) 
      + \ln\left( \frac{\mu^2}{m_q^2}\right)- 1\right],
\label{HardcontribNC}
\eqa
where $\mu$ is the factorization scale.
The energy of the emitted gluon is
\bq
p^0_5 = \frac{\sqrt{\hs}}{2} \, (1-z),
\label{p50}
\eq
where  ~$z = \ds \frac{s'}{\hs} $, ~$ \hs = - 2 (p_1.p_2) $,
$p_1$ and $p_2$ are momenta of the incoming quarks 
and  $s'= - (p_3+p_4)^2$ is the invariant mass of the outgoing 
leptons\footnote{We use the $(-,+,+,+)$ metrics, $p=(p^0,\vec{p})$.};
\bq
P_{qq}(z) =C_F\, \frac{1+z^2}{1-z}
\eq
is the leading order (LO) quark-quark splitting function. 

In the charged current case we have the limits of integration over 
variable $z$: $\ds z_{max} = 1 - \frac{2\, \bar \omega}{\sqrt{\hs}}$, 
where $\bar \omega \ll \sqrt{\hs}$  and $\ds z_{min}= \frac{\ml^2}{\hs}$,
$\ml$ being the mass of charged leptons. 
The auxiliary parameter $\bar \omega$ is the maximal energy of a soft
gluon in the c.m.s. of the incoming partons.
In the neutral current case we have correspondingly the same
 $ z_{max}$ and $\ds z_{min}= \frac{4 \ml^2}{\hs}$.

The cross sections in the Born approximation read
\bqa
{\hat \sigma}^{\rm CC}_0(\hs) &=& 
{\mid V_{ud}\mid}^2 \frac{{G^2_{\sss F}}}{18 \pi} \,\,
       \frac{\mw^4 \, {\hs} }{{\mid {\hat s-\mwt}\mid}^2} \,\,
\left(1 -\frac{3 \ml^2}{2 {\hs}} + \frac{m^6_\ell}{2 {\hs}^2} \right), \\
{\hat \sigma}^{\rm NC}_0(\hs) &=&
 \frac{4\, \pi \, \alpha^2}{3 \hs}\, \beta(\hs,\ml^2) \left[
\frac{1}{3}\, \left(1-\frac{\ml^2}{\hs} \right) V_0(\hs) 
        +\frac{ \ml^2}{\hs} V_a(\hs)\right],    
\label{BornNC}
\eqa
where
$ \hs = - (p_1+p_2)^2$, $p_1$ and $p_2$ are 4-momenta of the initial quarks;
 $\mwt = \mw^2-i\mw\Gamma_{\sss W}$;
$\ds \beta(\hs,\ml^2)=\sqrt{1-\frac{4 \ml^2}{\hs}}$. 
 Here we denoted
\bqa
V_0(\hs)&=& Q^2_q Q^2_\ell 
+ 2\,Q_q Q_\ell \, \mid \chi_{\sss Z}(\hs)\mid \,v_q \,v_\ell 
+ {\mid \chi_{\sss Z}(\hs) \mid}^2
          \left( v^2_q+{I^{(3)}_q}^2\right)\left( v^2_\ell+{I^{(3)}_\ell}^2\right),
\nll
V_a(\hs) &=&V_0(\hs)-2{\mid \chi_{\sss Z}(\hs)\mid}^2 
\left( v^2_q+{I^{(3)}_q}^2\right)\left({I^{(3)}_\ell}\right)^2,
\nll && v_q = I^{(3)}_q - 2 Q_q \sin^2\theta_W, \qquad  \quad
   v_\ell = I^{(3)}_\ell - 2 Q_\ell \sin^2\theta_W.
\label{V0Vavf}
\eqa
The $Z/\gamma$ propagator ratio $\chi_{\sss Z}(\hs)$
with $\hs$--dependent or constant $Z$-width is 
\bqa
 \chi_{\sss Z}(\hs)=\frac{\hs}{s - \mz^2 + i \hs \ds \frac{\Gamma_{\sss Z}}{\mz}}
 \,\, \frac{1}{4 \sin^2\theta_W \cos^2\theta_W}.
\label{chi}
\eqa 
We see in Eqs.~(\ref{HardcontribCC}) and (\ref{HardcontribNC}) that the 
collinear singularities appear as quark mass  singularities.

\subsection{Hard gluon contribution in \MSbar  scheme with massless quarks}
 
Because the collinear singularities calculated in the \MSbar scheme
with massless quarks are already included in PDF, we have
to find  which terms in our massive quarks treatment of the cross sections have 
to be subtracted to avoid the double counting. Therefore we calculate the same 
cross sections  in the \MSbar  scheme following the well known
\cite{AlElMar1979} manner of working with massless quarks and 
compare with our results.

Calculating the three particle phase space element of the hard gluon 
emission in the $n$-dimensional phase space we used a cascade in two steps:
\bqa
 d { \Phi^{(3)}}= \frac{d s'}{2 \pi}\,  d{\Phi^{(2)}_1} d{\Phi^{(2)}_2}.
\label{FS} 
\eqa
  For the first step of the charged current process
${\bar d}(p_1)+ u(p_2) \to W^*(Q')+g(p_5)$  we  obtained
the same formula as in ~\cite{AlElMar1979}:
\bqa
{ \Phi^{(2)}_1}=\frac{1}{8 \pi}\,
\left(\frac{4 \pi \,\mu^2}{s'} \right)^{\varepsilon} \,
           \frac{1}{\Gamma(1-\varepsilon)}
 \,  z^{\varepsilon}\, (1-z)^{1- 2 \varepsilon}\,\,
\int_0^1 d y \,\,{y}^{-\varepsilon} \, (1-y )^{-\varepsilon},
\label{PS1CCn}
\eqa
where we introduced in addition to $z$ the variable $y$
\bq
 y = \frac{1 + \cos(\theta_g)}{2}.
\label{variably}
\eq
Here $\theta_g$ is an angle  between vectors $\vec p_1$ and $\vec p_5$,  
the angle of the emitted gluon.

The phase space element of the second step 
$  W^*(Q') \to {\nu_\ell}(p_3) + {\ell^+}(p_4)$ of the cascade  is:
\bq
{ \Phi^{(2)}_2}=\frac{1}{16 \pi^2}\,
\left(\frac{4 \pi \,\mu^2}{s'} \right)^{\varepsilon} \, 
\frac{1}{\Gamma(1-\varepsilon)}
 \,  \left(1- \frac{\ml^2}{s'}\right)^{1- 2 \varepsilon}\,\,
\int_0^1 d y_R \,\,{y_R}^{-\varepsilon} \, (1-y_R )^{-\varepsilon}
\, \int_0^{2 \pi} d \varphi_R,
\label{PS2CCn}
\eq
where 
\bq
 y^R = \frac{1 + \cos(\theta^R)}{2},
\label{variablyR}
\eq
and $\theta^R$ is an angle between the charged lepton and gluon in the rest
frame of the outgoing leptons.

Having in mind that the cross section of the charged current process
in Born approximation has the form 
\bqa
{\hat \sigma}^{\rm CC}_0(\hs, \varepsilon) &=&
{\mid V_{ud}\mid}^2 \,
       \frac{{G^2_{\sss F}} \mw^4  \hs }{6 \pi {\mid {\hs-\mwt}\mid}^2} 
 \left(\frac{4 \pi \,\mu^2}{\hs} \right)^{\varepsilon}
 \frac{1}{\Gamma(1-\varepsilon)}
\left(1 -\frac{\ml^2}{\hs} \right)^{2-2 \varepsilon}
\int_0^1 d y_0 \,{y_0}^{-\varepsilon} (1-y_0 )^{-\varepsilon}  \nll
 &&\left[y_0 - \left(1 -\frac{\ml^2}{\hs} \right) y_0 (1-y_0 ) 
-\frac{1}{2} \varepsilon \right],
\label{n_BornCC}
\eqa 
we  obtained a factorized expression 
 of the hard gluon NLO correction to the charged current process.
 \bqa
 {\hat \sigma}^{\rm CC}_{\rm Hard}(\varepsilon) & = &
 \frac{\als}{2 \pi} \, C_F \int_{z_{min}}^{z_{max}}  d z \,\,
{\hat \sigma}^{\rm CC}_0(z \hs,\varepsilon) \,
\left(\frac{4 \pi \mu^2}{z \hs} \right)^{\varepsilon} \,
      \frac{1}{\Gamma(1-\varepsilon)}
 \,  z^{\varepsilon}\, (1-z)^{- 2 \varepsilon}\,\,
\int_0^1 d y \,\,{y}^{-\varepsilon} \, (1-y )^{-\varepsilon} \nll 
 && \left[\frac{1}{y\,(1-y)}\,\left(\frac{1}{1-z} -1 +\frac{1}{2} \,(1-z)\right)
+(1-z)\,\left(- 1 - \frac{\varepsilon}{2\,y\,(1-y)}\right)
\right],
\label{HardCCnyz}
\eqa
Integration over $y$ gives
\bqa
 {\hat \sigma}^{\rm CC}_{\rm Hard}(\varepsilon) = 
 \frac{\als}{2 \pi} \int_{z_{min}}^{z_{max}}  d z
 \,{\hat \sigma}^{\rm CC}_0(z \hs,\varepsilon) \,
\left(\frac{4 \pi \mu^2}{z \hs} \right)^{\varepsilon} \,
      \frac{\Gamma(1-\varepsilon)}{\Gamma(1-2 \varepsilon)}
 \,  z^{\varepsilon}\, (1-z)^{- 2 \varepsilon}\,P_{qq}(z)\,
     \left( - \frac{2}{\varepsilon} \right).
\label{HardCCnz}
\eqa
One can see that the collinear divergence appears here
as a pole $\ds \frac{1}{\varepsilon}$. To compare this expression 
with the analogous expression (\ref{HardcontribCC}) where the  
collinear divergence manifests itself in the form of logarithms
$\ds \ln\left( \frac{\hs}{\mup^2}\right)$,
$\ds \ln\left( \frac{\hs}{\mdn^2}\right)$ one have to take
the limit $\varepsilon \to 0$ . Then  one obtains an expression
\bq
  {\hat \sigma}^{\rm CC}_{\rm Hard} = \frac{\als}{2 \pi}
\, \int_{z_{min}}^{z_{max}}  d z \,
 \,{\hat \sigma}^{\rm CC}_0(z \hs) \,
\,P_{qq}(z)\, \left(-\frac{2}{\bar \varepsilon} 
+ 2 \ln\left( \frac{\hs}{\mu^2}\right)
+ 4 \ln(1-z) \right)
\label{HardCCeps}
\eq
to be compared with the corresponding expression (\ref{HardcontribCC}).
We see which terms in the expression (\ref{HardcontribCC}) correspond to
 the collinear divergent term $\ds -\frac{1}{\bar \varepsilon} $
which in \MSbar scheme has to be subtracted from the hard gluon
 contribution
to the considered process because it is already included into PDF.

So, on the quark-parton level we have subtract from 
$ {\hat \sigma}^{\rm CC}_{\rm Hard}$ (\ref{HardcontribCC}),
 the following expression: 
\bqa
  {\hat \sigma}^{\rm CC}_{H_{\rm Subtr}}(\mu^2) = \frac{\als}{2 \pi}
\, \int_{z_{min}}^{z_{max}}  d z \,
 \,{\hat \sigma}^{\rm CC}_0(z \hs) \,\, P_{qq}(z) \,\,
\left[\ln\left(\frac{\mu^2}{\mup^2}\right)
+\ln\left( \frac{\mu^2}{\mdn^2}\right) -2 - 4\ln(1-z)\right].
\label{subtrHardCC}
\eqa
Factorization properties and general relations between amplitudes
with massive and massless partons can be found in Ref.~\cite{Moch:2007pj}.

For the neutral current process calculating the three particle phase 
space element (\ref{FS}) of the hard gluon emission in the $n$-dimensional 
phase space  we obtain  the same result (\ref{PS1CCn}) for the first step 
${\bar q}(p_1)+ q(p_2) \to \{\gamma, Z\}^*(Q')+g(p_5)$.
The phase space element of the second step 
$\{\gamma, Z\}^*(Q') \to {\ell^-}(p_3) + {\ell^+}(p_4)$ of the cascade  is
similar to (\ref{PS2CCn}):
\bq
{ \Phi^{(2)}_2}=\frac{1}{16 \pi^2}\,
\left(\frac{4 \pi \,\mu^2}{s'} \right)^{\varepsilon} \, 
\frac{1}{\Gamma(1-\varepsilon)}
 \, \beta^{1- 2 \varepsilon}(s',\ml) \,\,
\int_0^1 d y_R \,\,{y_R}^{-\varepsilon} \, (1-y_R )^{-\varepsilon}
\, \int_0^{2 \pi} d \varphi_R.
\label{PS2NCn}
\eq

Analogously, we  obtained a factorized expression 
for the hard gluon NLO correction to the neutral current process. 
It has exactly the same structure as Eq.~(\ref{HardCCnyz}), 
but the cross section in Born approximation
${\hat \sigma}^{\rm NC}_0(z \hs,\varepsilon)$ 
 of the neutral current process has a different form: 
\bqa
{\hat \sigma}^{\rm NC}_0(\hs, \varepsilon) &=&
\frac{4\, \pi \, \alpha^2}{3 \hs}\, \beta^{1-2 \varepsilon}(\hs,\ml^2) 
 \left(\frac{4 \pi \,\mu^2}{\hs} \right)^{\varepsilon}
 \frac{1}{\Gamma(1-\varepsilon)}\,\, 
\int_0^1 d y_0 \,\,{y_0}^{-\varepsilon} \, (1-y_0 )^{-\varepsilon}  \nll
 &&\Biggl[V_0(\hs) \left( - \beta^2(\hs,\ml^2)\,  y_0 (1-y_0) 
+ (1-\varepsilon) \left(\frac{1}{2} - \frac{\ml^2}{\hs} \right) \right) \nll
&&+  V_a(\hs)\,(1-\varepsilon) \,\frac{ \ml^2}{\hs}
+ A_0(\hs) (1-\varepsilon)(1- 2\varepsilon)\,\beta(\hs,\ml^2)(\frac{1}{2} - y_0)
 \Biggr],
\label{n_BornNC}
\eqa 
where 
\bq
A_0(\hs)= \qquad  \quad  2\,Q_q Q_\ell \,
   \mid \chi_{\sss Z}(\hs)\mid \,I^{(3)}_q \,I^{(3)}_\ell 
+ {\mid \chi_{\sss Z}(\hs) \mid}^2\, 4\,v_q \, v_\ell\,I^{(3)}_q \,I^{(3)}_\ell.
\eq 
Integration over $y$ gives for ${\hat \sigma}^{\rm NC}_{\rm Hard}(\hs, \varepsilon)$
the same result as (\ref{HardCCnz}). So, in the limit $\varepsilon \to 0$ 
 we come to the expression almost the same as  (\ref{HardCCeps}),
\bq
  {\hat \sigma}^{\rm NC}_{\rm Hard} = \frac{\als}{2\pi}
\, \int_{z_{min}}^{z_{max}}  d z \,
 \,{\hat \sigma}^{\rm NC}_0(z \hs) \,
\,P_{qq}(z)\, \left(-\frac{2}{\bar \varepsilon} 
+ 2\ln\left( \frac{\hs}{\mu^2}\right)
+ 4\ln(1-z) \right).
\label{HardNCeps}
\eq

Comparison of this expression where the collinear divergence appears 
as a pole $\ds \frac{1}{\varepsilon}$, with (\ref{HardcontribNC}) where the  
collinear divergence manifests itself in the form of logarithms permits us
to find what expression one has to subtract from  
$ {\hat \sigma}^{\rm NC}_{\rm Hard}$ (\ref{HardcontribNC}), namely
\bqa
  {\hat \sigma}^{\rm NC}_{H_{\rm Subtr}}(\mu^2) = \frac{\als}{\pi}
\, \int_{z_{min}}^{z_{max}}  d z \,
 \,{\hat \sigma}^{\rm NC}_0(z \hs) \,\, P_{qq}(z) \,\,
\left[\ln\left(\frac{\mu^2}{m_q^2}\right) -1 - 2\ln(1-z)\right], 
\label{subtrHardNC}
\eqa
where $m_q$ is the mass of the pair quark and antiquark coming from the 
both protons.

\subsection{Virtual and soft gluon contribution}

Working in our massive quark treatment we obtained for the sum of
virtual and soft gluon contributions an expression free from infrared 
divergences. In the case of charged current processes we have
\bqa
   {\hat \sigma}^{\rm CC}_{\rm Virt}
+  {\hat \sigma}^{\rm CC}_{\rm Soft} = 
 \frac{\als}{2 \pi}   \, C_F \,  {\hat \sigma}^{\rm CC}_0(\hs)
 \,\Bigg\{ \left( \frac{3}{2} +
\ln\left(\frac{4 {\bar \omega}^2}{\hat s} \right) \right) 
\left[ \ln\left( \frac{\hat s}{\mup^2}\right) 
     + \ln\left( \frac{\hat s}{\mdn^2}\right)- 2\right] 
                                      -1 - \frac{\pi^2}{3}  \Bigg\}.
\label{SoftVirtCC}
\eqa
Correspondingly, in the case of neutral current processes we have
\bqa
   {\hat \sigma}^{\rm NC}_{\rm Virt}
+  {\hat \sigma}^{\rm NC}_{\rm Soft} =  \frac{\als}{\pi}\, C_F\,
{\hat \sigma}^{\rm NC}_0(\hs) \,   \Bigg\{ \left(  \frac{3}{2} +
        \ln\left(\frac{4 {\bar \omega}^2}{\hs}\right) \right) \, 
  \left[\ln\left(\frac{\hs}{m_q^2}\right) -1\right] 
                           - \frac{1}{2} - \frac{\pi^2}{6} \Bigg\}.
\label{SoftVirtNC}
\eqa

One can find the collinear divergent expressions to be subtracted from these 
virtual and soft gluon contributions because they are  already included 
into PDF, taking the corresponding expressions which one has subtract 
from the hard gluon contributions to the considered processes.  But one has to
take them with opposite sign, to substitute the argument of Born cross 
sections taken  $z = 1$, namely 
\bqa
{\hat \sigma}^{\rm CC}_0(z \hs) \Longrightarrow {\hat \sigma}^{\rm CC}_0(\hs)
 \qquad \mbox{and } \qquad 
{\hat \sigma}^{\rm NC}_0(z \hs) \Longrightarrow {\hat \sigma}^{\rm NC}_0(\hs)
\eqa
and to integrate over $z$  from 0 to $z_{max}$. In this way we obtained the
expressions to be subtracted from  virtual and soft gluon contributions.

For charged current processes collinear divergent subtraction   is:
\bqa
  {\hat \sigma}^{\rm CC}_{SV_{\rm Subtr}}(\mu^2) = - \frac{\als}{2 \pi}\,\,
{\hat \sigma}^{\rm CC}_0(\hs) \, \int_0^{z_{max}}  d z \,\, P_{qq}(z) \,\,
\left[\ln\left(\frac{\mu^2}{\mup^2}\right)
+\ln\left( \frac{\mu^2}{\mdn^2}\right) -2 - 4\ln(1-z)\right].
\label{subtrVirtSoftCC}
\eqa
And for neutral current processes it is:
\bqa
  {\hat \sigma}^{\rm NC}_{SV_{\rm Subtr}}(\mu^2) = - \frac{\als}{\pi}\,\,
{\hat \sigma}^{\rm NC}_0(\hs)\, \int_0^{z_{max}}  d z \,\, P_{qq}(z) \,\,
\left[\ln\left(\frac{\mu^2}{m_q^2}\right) -1 - 2\ln(1-z)\right], 
\label{subtrVirtSoftNC}
\eqa
Integration over $z$ gives the expressions to be found, namely,
 first - a subtraction for the charged current contribution:
\bqa
  {\hat \sigma}^{\rm CC}_{SV_{\rm Subtr}}(\mu^2) &=&  \frac{\als }{2 \pi}\,C_F\,
{\hat \sigma}^{\rm CC}_0(\hs) \Biggl\{ \left( \frac{3}{2} +
\ln\left(\frac{4 {\bar \omega}^2}{\hat s} \right) \right)
\left[ \ln\left( \frac{\mu^2}{\mup^2}\right) 
     + \ln\left( \frac{\mu^2}{\mdn^2}\right)- 2\right] \nll
 && \hspace{7.5cm}
 + 7 - \ln^2\left(\frac{4 {\bar \omega}^2}{\hat s} \right) \Biggr\}
\label{subtrVirtSoftCCint}
\eqa
and second - a subtraction for the neutral current contribution:
\bqa
 {\hat \sigma}^{\rm NC}_{SV_{\rm Subtr}}(\mu^2) = \frac{\als}{\pi}\,C_F\,
{\hat \sigma}^{\rm NC}_0(\hs) \left\{ \left( \frac{3}{2} +
\ln\left(\frac{4 {\bar \omega}^2}{\hat s} \right) \right)
\left[\ln\left(\frac{\mu^2}{m_q^2}\right) -1 \right] + \frac{7}{2} 
- \frac{1}{2}\ln^2\left(\frac{4 {\bar \omega}^2}{\hat s} \right) \right\}. 
\label{subtrVirtSoftNCint}
\eqa
One can see that subtractions (\ref{subtrVirtSoftCCint}) and 
(\ref{subtrVirtSoftNCint}) really subtract the collinear divergent terms
in the expressions of virtual and soft gluon contributions
(\ref{SoftVirtCC}) and (\ref{SoftVirtNC}), correspondingly.

\section{Quark--gluon  Drell--Yan processes }
\subsection{Charged current quark--gluon processes }

\begin{figure}[ht]
\[
\begin{picture}(400,86)(80,0)
  \Photon(100,43)(175,43){3}{10}
    \Vertex(175,43){2}
  \ArrowLine(200,86)(175,43)
  \ArrowLine(175,43)(200,0)
  \ArrowLine(75,0)(100,43)
    \Vertex(100,43){2}
  \ArrowLine(100,43)(112,64)
    \Vertex(112,64){2}
  \ArrowLine(112,64)(125,86)
    \Gluon(68,86)(112,64){3}{6}
\Text(125,78)[lb]  { $d(p_1)$}
\Text(79,9)[lt]  { $u(p_2)$}
\Text(135,25)[bc] { $(W^+,\phi^+)$}
\Text(200,78)[lb]{ $\ell^+(p_4)$}
\Text(202,10)[lt]{ $\nu_{\ell}(p_3)$}
\Text(50,75)[lt]{ $g(p_5)$}
  \Photon(375,43)(450,43){3}{10}
    \Vertex(450,43){2}
  \ArrowLine(475,86)(450,43)
  \ArrowLine(450,43)(475,0)
  \ArrowLine(350,0)(362,21)
    \Vertex(362,21){2}
  \ArrowLine(362,21)(375,43)
    \Vertex(375,43){2}
  \ArrowLine(375,43)(400,86)
  \Gluon(319,42)(362,21){3}{6}
\Text(400,78)[lb]  { $d(p_1)$}
\Text(354,9)[lt]  { $u(p_2)$    }
\Text(410,25)[bc] { $(W^+,\phi^+)$  }
\Text(475,78)[lb]{ $\ell^+(p_4)$}
\Text(477,10)[lt]{ $\nu_{\ell}(p_3)$      }
\Text(320,60)[lt]{ $g(p_5)$}
\end{picture}
\]
\caption[Gluon inverse bremsstrahlung] 
        {Charged current diagrams with coming gluon.}
\label{GluonDYCC}
\end{figure}
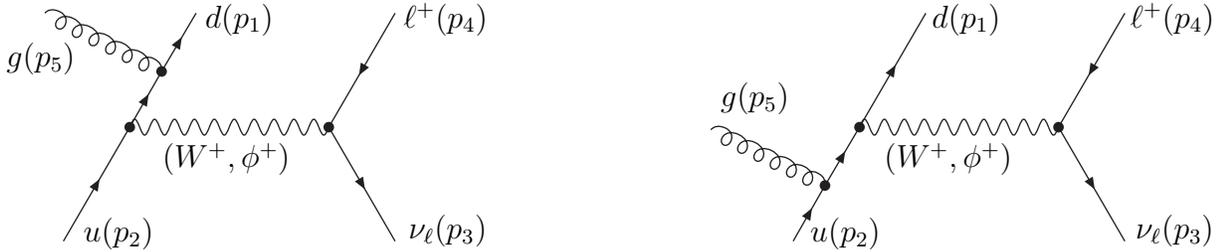

In the framework of the  Drell--Yan 
process ${p}\, {p} \to W \to \ell\, {\nu}_\ell$
we have to take into account the presence of gluons in the protons.
Therefore we  consider on the quark-parton level the attendant
processes with incoming gluon
~$u(p_2) +  g(p_5)  \to d(p_1) + \nu_\ell(p_3)+ \ell^+(p_4) $, 
see the Feynman diagrams  on the Fig.\ref{GluonDYCC},
and  ~$ \bar d(p_1) +  g(p_5) \to \bar u(p_2) + \nu_\ell(p_3) + \ell^+(p_4) $
(similar diagrams). 

The contributions of these processes do not contain infrared divergences
but have  quark mass singularities.  

The three particle phase space element of the process
 ~$u~ g~  \to d~ \ell^+ ~\nu_\ell$
 can be treated as a cascade  analogously to (\ref{FS})
 but here gluon $g$ is incoming and $d $ - quark is outgoing.
The phase space element of the first step 
 $g(p_5)+ u(p_2) \to W^*(Q') + d(p_1)$ of the cascade is the same as in Eq.~(\ref{PS1CCn}).
Variable $z $ has the same meaning, $z = \ds \frac{s'}{\hs}$, 
and  $s'= - (p_3+p_4)^2$ is the invariant mass of outgoing leptons,
but here ~$\hs = - 2 (p_2.p_5)$.
Variable $y$ has the same form (\ref{variably}) and $\theta_g$ is 
the angle  between vectors $\vec p_1$ and $\vec p_5$,
but now $\vec p_5$ is the momentum of the incoming gluon and $\vec p_1$ is 
the momentum of the outgoing quark.

The phase space of the second step 
$  W^*(Q') \to {\nu_\ell}(p_3) + {\ell^+}(p_4)$ of the cascade, obviously, 
has the same form (\ref{PS2CCn}).

In our massive quarks treatment in the 4-dimensional phase space  we obtain 
the cross section of the process 
~$u  ~g  \to d  ~\nu_\ell ~\ell^+ $
 in the form factorized to the Born cross 
section of the corresponding quark-antiquark process
~$u ~\bar d \to \nu_\ell ~\ell^+ $:
\bqa
   {\hat \sigma}^{\rm CC}_{ug} = \frac{\als}{2 \pi}
 \int_{z_{min}}^{z_{max}}  d z  \,\, {\hat \sigma}^{\rm CC}_0(z \hs) \,\, 
    \left\{  P_{qg}(z) \,\,
\left[\ln\left( \frac{\hs}{\mdn^2}\right) + 2\, \ln(1-z) - \frac{7}{4} \right]
  +\frac{9}{8} - \frac{1}{4} \, z  \right\}, 
\label{GluoncontribCCu}
\eqa
where 
\bq
P_{qg}(z) = T_f\,\left[z^2 + (1-z)^2 \right]
\eq
is the quark-gluon splitting function and  $T_f=\frac{1}{2}$. We
neglected the quark masses except of in the logarithm, where we have a 
mass of  the outgoing quark, namely we face a mass singularity.
Integration over $z$ here is up to
 $\ds z_{max}= \left(1-\frac{\mdn}{\sqrt{\hs}} \right)^2 \approx 1$.

In the \MSbar  scheme with massless quarks working in
the $n$-dimensional phase space we obtain a factorized expression:
 \bqa
   {\hat \sigma}^{\rm CC}_{ug}(\varepsilon) &=& \frac{3 \als}{8 \pi}\, 
 \frac{C_F}{(1-\varepsilon)} \, \int_{z_{min}}^{z_{max}}  d z \,
 {\hat \sigma}^{\rm CC}_0(z \hs,\varepsilon) \,  
\left(\frac{4 \pi \mu^2}{z \hs} \right)^{\varepsilon} \,
      \frac{ z^{\varepsilon}\, (1-z)^{- 2 \varepsilon}}{\Gamma(1-\varepsilon)}
\int_0^1 d y \,\,{y}^{-\varepsilon} \, (1-y )^{-\varepsilon} \nll  
&\times& \left\{ \frac{1}{y}\left(z^2-z+\frac{1}{2}\,(1-\varepsilon) \right)
+ z - z^2 + (1-z)\,\varepsilon + \frac{1}{2}\,y\,(1-z)^2\,(1-\varepsilon)
    \right\}. 
\label{GluonQuark0CCn}
\eqa

We divided here  by $1-\varepsilon$ because of the number of 
 the gluon spin projections in the $n$-dimensional phase space is
$n - 2 =  2 (1 - \varepsilon)$.  After integration over $y$ we have 
\bqa
  {\hat \sigma}^{\rm CC}_{ug}(\varepsilon) &=& 
\frac{ \als}{2 \pi} \,\,   \int_{z_{min}}^{z_{max}}  d z \,
{\hat \sigma}^{\rm CC}_0(z \hs,\varepsilon) \,
\left(\frac{4 \pi \mu^2}{z \hs} \right)^{\varepsilon} \,
      \frac{\Gamma(1-\varepsilon)}{\Gamma(1 - 2 \varepsilon)}
 \, \frac{z^{\varepsilon}\, (1-z)^{- 2 \varepsilon}}{1-\varepsilon} \nll               
 &\times& \biggl\{ P_{qg}(z) \,\, \left[ - \frac{1}{\varepsilon} 
     - \frac{3}{4} -\frac{1}{4}\,\varepsilon \right]
  +\frac{9}{8} - \frac{1}{4} \, z 
   +\varepsilon\,\left( \frac{7}{8} - \frac{3}{4}\,z\right) 
                \biggr\}.
\label{GluonQuark1CCn}
\eqa
In this expression the collinear divergence 
appears as a pole $ \ds \frac{1}{\varepsilon}$.
It has to be compared with the analogous expression (\ref{GluoncontribCCu}) 
where the collinear divergence appears as a mass singularity.
In the limit $\varepsilon \to 0$ we have
\bq
-\frac{1}{\varepsilon}\,\left(\frac{4 \pi \mu^2}{z \hs} \right)^{\varepsilon} \,
      \frac{\Gamma(1-\varepsilon)}{\Gamma(1-2 \varepsilon)}
 \, \frac{z^{\varepsilon}\, (1-z)^{- 2 \varepsilon}}{1-\varepsilon}= 
-\frac{1}{\bar \varepsilon} -1 + \ln\left( \frac{\hs}{\mu^2}\right)
+ 2 \ln(1-z) + {\mathcal O}(\varepsilon).
\label{limiteps}
\eq
In this way we obtain
\bqa
  {\hat \sigma}^{\rm CC}_{ug}(\varepsilon) = \frac{ \als}{2 \pi}\,
  \int_{z_{min}}^{z_{max}}  d z \,
 {\hat \sigma}^{\rm CC}_0(z \hs) \,
\biggl\{  P_{qg}(z)  \left[ - \frac{1}{\bar \varepsilon} 
+ \ln\left( \frac{\hs}{\mu^2}\right) + 2 \ln(1-z)   - \frac{7}{4}\right]
  +\frac{9}{8} - \frac{1}{4} \, z \biggr\}.
\label{GluonQuark2CCn}
\eqa
Comparing with (\ref{GluoncontribCCu}) we see that the collinear divergent term 
which in \MSbar scheme has to be subtracted from the cross section
of the considered process (because it is already included into PDF 
 of  gluons) in our treatment has the form
\bqa
 {\hat \sigma}^{\rm CC}_{ug_{Subtr}} = \frac{\als}{2 \pi} 
\, \int_{z_{min}}^1 {d z} \,\,
 {\hat \sigma}^{\rm CC}_0(z \hs) \,\, 
  P_{qg}(z)  \left[ \ln\left( \frac{\mu^2}{\mdn^2}\right) \right].
\label{deltaGluonCCu}
\eqa
Here $\mdn$ is the mass of the outgoing quark.

We have the same situation with the process  
~$g ~\bar d \to \bar u ~\nu_\ell ~\ell^+$.
Analogously  we obtain that from the cross section of this process we
have to subtract  the corresponding expression:
\bqa
 {\hat \sigma}^{\rm CC}_{dg_{Subtr}} = \frac{\als}{2 \pi} 
\, \int_{z_{min}}^1 {d z} \,\,
 {\hat \sigma}^{\rm CC}_0(z \hs) \,\, 
  P_{qg}(z)  \left[ \ln\left( \frac{\mu^2}{\mup^2}\right)  \right].
\label{deltaGluonCCd}
\eqa
In this process anti-up quark is outgoing and it can move collinearly with
the gluon, so the logarithm is from the mass  $\mup$ of the $u$-quark.

\subsection{Neutral current quark-gluon processes }

In the case of  the DY processes 
$p\, p \to [A, Z] \to \ell^-\, \ell^+ $
we  consider on the quark-parton level the additional process 
~$q(p_2) +  g(p_5)  \to q(p_1) + \ell^-(p_3) + \ell^+(p_4) $ (corresponding
Feynman diagrams are similar to the Fig.\ref{GluonDYCC}) and  process
~$ \bar q(p_1) +  g(p_5)  \to \bar q(p_2) + \ell^-(p_3)+\ell^+(p_4) $ 
with an incoming gluon.
The contributions of these processes also do not contain infrared 
divergences but have  quark mass singularities. 
We considered these processes analogously as the charged   
current quark--gluon processes. Difference is only in the Born cross 
section to which they are factorized. 
 
In our massive quarks treatment in the 4-dimensional phase space  
the cross section  of both kind of neutral current quark--gluon processes
have equal form:
\bqa
   {\hat \sigma}^{\rm NC}_{qg} = \frac{\als}{2 \pi} \,
 \int_{z_{min}}^{z_{max}}  d z \,
{\hat \sigma}^{\rm NC}_0(z \hs) \,     \left\{  P_{qg}(z) \,\,
\left[\ln\left( \frac{\hs}{m_q^2}\right)-1 + 2\, \ln(1-z) - \frac{3}{4} \right]
  +\frac{9}{8} - \frac{1}{4} \, z  \right\}, 
\label{GluoncontribNCq}
\eqa
where $m_q$ is the mass of incoming as well as outgoing quark (or anti-quark).
The upper limit of integration over $z$ is the same, namely
 $\ds z_{max}= \left(1-\frac{m^2_q}{\sqrt{\hs}} \right)^2 \approx 1$.
Here we also neglected the quark masses except of under the logarithm.

In the \MSbar  scheme with massless quarks working in
the $n$-dimensional phase space we obtain the same expressions 
(\ref{GluonQuark0CCn}), (\ref{GluonQuark1CCn}), (\ref{GluonQuark2CCn}). as
those of charged current quark--gluon processes. 

Analogously comparing gave us that the collinear divergent term 
which in \MSbar scheme has to be subtracted from the cross section
of the both considered processes, because is already included into PDF 
 of  gluons, in our treatment has a form
\bqa
 {\hat \sigma}^{\rm NC}_{qg_{Subtr}} = \frac{\als}{2 \pi} 
\, \int_{z_{min}}^1 {d z} \,\,
 {\hat \sigma}^{\rm NC}_0(z \hs) \,\, 
  P_{qg}(z)  \left[ \ln\left( \frac{\mu^2}{m^2_q}\right) \right].
\label{deltaGluonNCq}
\eqa

\section{Numerical calculations on hadronic level}

\subsection{Hadronic level kinematics} 

\begin{figure}[!ht]
\[
\begin{picture}(400,90)(0,0)
  \LongArrow(0,50)(57,50)
  \LongArrow(120,50)(63,50)
  \LongArrow(125,50)(140,50)
  \LongArrow(63,52)(100,90)
  \LongArrow(57,48)(20,10)
  \CArc(60,50)(25,180,225)
  \CArc(60,50)(25,0,45)
\Text(90,55)[lb] {$\theta_q$}
\Text(22,32)[lb] {$\theta_q$}
\Text(135,40)[lb]{$z$}
\Text(105,44)[lt]  { ${\vec p_2}$ }
\Text(5,55)[lb]  { ${\vec p_1}$}
\Text(80,85)[lb] {${\vec p_3}$}
\Text(35,10)[lb] {${\vec p_4}$}
\Text(60,0)[lt] {a)}
  \LongArrow(200,50)(257,50)
  \LongArrow(320,50)(263,50)
  \LongArrow(325,50)(340,50)
  \LongArrow(263,52)(310,80)
  \LongArrow(257,48)(210,20)
  \CArc(260,50)(25,0,30)
  \CArc(260,50)(25,180,210)
\Text(290,55)[lb] {$\theta_N$}
\Text(220,37)[lb] {$\theta_N$}
\Text(335,40)[lb]{$z$}
\Text(305,44)[lt]  { ${\vec {p_2}_N}$ }
\Text(205,55)[lb]  { ${\vec {p_1}_N}$}
\Text(295,80)[lb] {${\vec p_3}$}
\Text(225,15)[lb] {${\vec p_4}$}
\Text(260,0)[lt] {b)}
\end{picture}
\]
\vspace{-.5cm}
\caption[Systems]{ a) Quark c.m.s.  ~b) Proton c.m.s. (Born approximation). }
\label{DYkinem}
\end{figure}
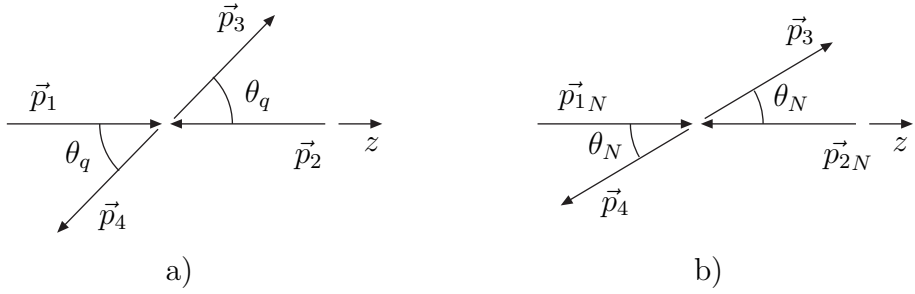

 In the c.m.s. of the quark-quark or quark-gluon pair   
 (see Fig.{\ref{DYkinem}a}) we have (neglecting masses of quarks):
\bqa
p^0_1=p^0_2=\mid{\vec p_1}\mid=\mid{\vec p_2}\mid=\frac{\sqrt{\hs}}{2},
\qquad  \qquad  \vec p_2 = - \vec p_1, \qquad \qquad \hs = - 2 p_1 \cdot p_2.
\label{BornKinemCC}
\eqa

In the c.m.s. of protons $p, p$ (see Fig.{\ref{DYkinem}b}) we have 
\bqa
&&{p_1^0}_N ={p_2^0}_N=\mid{\vec {p_1}_N}\mid = \mid{\vec {p_2}_N}\mid=
         \frac{\sqrt{s}}{2}, \qquad \quad  \vec {p_2}_N = - \vec {p_1}_N,
\qquad \quad s = - 2 {p_1}_N \cdot {p_2}_N  \nll 
 &&    p_1 = x_1\,{p_1}_N, \quad  p_2 = x_2\,{p_2}_N,\qquad \quad
 \hs = - 2 p_1 \cdot p_2 = - 2 {p_1}_N \cdot {p_2}_N\, x_1 x_2 = s\, x_1  x_2.
\label{x1,x2}
\eqa
Equations of transition from quarks c.m.s to protons c.m.s. for any
momentum have the form
\bq
{\hat Q}^0 = \ds \gamma \left( Q^0_N - \beta \, Q^z_N \right), \qquad 
\qquad {\hat Q}^z = \ds \gamma \left( Q^z_N - \beta \, Q^0_N \right),
\eq
where 
\bq
  \gamma = \ds  \frac{x_1 + x_2}{2 \sqrt{x_1\, x_2}},  \qquad \qquad \qquad
  \gamma \beta = \ds  \frac{x_1 - x_2}{2 \sqrt{x_1\, x_2}}.
\eq

Replacing the scalar products 
\bq
p_1 \cdot p_3 = - {p_1}^0 \, {p_3}^0 \, \left(1-\cos{\theta_q}\right), \qquad
\qquad  p_2\cdot p_3 = -  {p_2}^0 \, {p_3}^0\, \left(1+\cos{\theta_q}\right),
\eq
where $\theta_q$ is an angle between the 3-momenta of the quark
 ${\vec p_1}$   and neutrino ${\vec p_3}$ ,
from quarks c.m.s. to protons c.m.s. 
\bq
p_1 \cdot p_3 = x_1\, {p_1}_N \cdot p_3 = 
 - x_1 \, {p_1^0}_N \, {p_3}^0 \, \left(1-\cos{\theta_N}\right), \qquad
p_2 \cdot p_3 = x_2\, {p_2}_N \cdot p_3 = - x_2\, {p_2^0}_N \, {p_3}^0 \,
       \left(1+\cos{\theta_N}\right),
\eq
and the angle $\theta_N$ is between
 3-momenta of the proton $\vec{p}_{1_N}$ and neutrino ${\vec p_3}$, after 
some algebra we come to the relation between 
 angles $\theta_q$ and  $\theta_N$
\bq
\cos{\theta_q}=\frac{x_2-x_1+(x_1+x_2)\cos{\theta_N}}
                    {x_1+x_2+(x_2-x_1)\cos{\theta_N}}.
\label{cosrelation}
\eq

The same connection we have for the angle $\theta^{14}_q$  between 
the quark momentum  $\vec{p}_{1}$ 
and the charged lepton momentum  ${\vec p_4}$ and for the angle  $\theta^{14}_N$
correspondingly between the proton momentum $\vec{p}_{1_N}$ and the 
charged lepton momentum  ${\vec p_4}$. We have only to take into account 
that for the massive charged lepton we have multiply $\cos{\theta_q}$
and correspondingly $\cos{\theta_N}$ by 
$\ds \frac{\mid \vec p_4\mid}{{p_4}^0}$, what is not equal to 1 for a
massive charged lepton. The same is valid for the neutral current case when
both final leptons are massive.

   Working with particle momenta components in the  protons c.m.s. we can
determine the parameters needed for the presentation of results on 
a hadronic level:
\begin{itemize}
\item
the transverse momenta of the charged lepton  and the missing
transverse momenta of the neutrino.
\bq
{p_4}_N^{\bot} = \sqrt{({p_4}_N^x)^2 +({p_4}_N^y)^2} \qquad \qquad
\mbox{and} \qquad \qquad  {p_3}_N^\bot = \sqrt{({p_3}_N^x)^2 +({p_3}_N^y)^2} .
\eq
\item  the rapidity of the charged lepton
\bq
\eta = \frac{1}{2} \ln{\frac{{p_4}_N^0+{p_4}_N^z}{{p_4}_N^0-{p_4}_N^z}}
\eq  
\item
the transverse invariant mass of the final leptons
\bq
M^{\bot} = \sqrt{2 {p_4}_N^{\bot} {p_3}_N^\bot 
        \left(1 - \cos{\varphi_N^{34}}\right)},
\eq
\end{itemize}
where $\varphi_N^{34}$ is an angle between the final leptons in a plane 
transversal to $z$ axis.

\subsection{Integration over variables $x_1$ and $x_2$}

We have to integrate over variables $x_1$ and $x_2$ the whole contribution
to the cross section of the DY charged current as well as 
neutral current processes. The leading order contribution is 
\bqa
{\sigma}^{\rm CC}_{LO} & = & \sum_{q_1q_2} \int_0^1 d x_1 f(x_1,\mu^2)
  \int_0^1 d x_2 f(x_2,\mu^2) \,  {\hat \sigma}^{\rm CC}_0( x_1 x_2\, s) ,  \nll 
{\sigma}^{\rm NC}_{LO} & = & \sum_{q} \int_0^1 d x_1 f(x_1,\mu^2)
  \int_0^1 d x_2 f(x_2,\mu^2)\, {\hat \sigma}^{\rm NC}_0( x_1 x_2\, s). 
\eqa   

In the next to leading order (NLO) we have to add  one-loop corrections:
 hard gluon contribution, virtual and soft gluon contribution and
corresponding subtractions of quark-antiquark processes and also the
contribution of quark-gluon processes with their subtractions. 
\bqa
{\sigma}^{\rm CC}_{NLO} & = & \sum_{q_1q_2} \int_0^1 d x_1\, f(x_1,\mu^2)
 \int_0^1 d x_2 \, f(x_2,\mu^2) \, 
\Bigg[ {\hat \sigma}^{\rm CC}_0( x_1 x_2\, s)    \nll
&+&      {\hat \sigma}^{\rm CC}_{\rm Hard}( x_1 x_2\, s)
       - {\hat \sigma}^{\rm CC}_{H_{\rm Subtr}}(\mu^2, x_1 x_2\, s)
       + {\hat \sigma}^{\rm CC}_{\rm SoftVirt}( x_1 x_2\, s)
       - {\hat \sigma}^{\rm CC}_{SV_{\rm Subtr}}(\mu^2, x_1 x_2\, s)\Bigg]\nll
&+&  \sum_{q_1 g}\int_0^1 d x_1\,  f(x_1,\mu^2)\int_0^1 d x_5 \, g(x_5,\mu^2) \,
\Bigg[
   {\hat \sigma}^{\rm CC}_{q_1g}( x_1 x_5\, s) 
 - {\hat \sigma}^{\rm CC}_{q_1g_{Subtr}}(\mu^2, x_1 x_5\, s)\Bigg]\nll
&+&  \sum_{q_2 g}\int_0^1 d x_2\,  f(x_2,\mu^2)\int_0^1 d x_5 \, g(x_5,\mu^2) \,
 + {\hat \sigma}^{\rm CC}_{q_2g}( x_2 x_5\, s) 
 - {\hat \sigma}^{\rm CC}_{q_2g_{Subtr}}(\mu^2, x_2 x_5\, s)
\Bigg].
\eqa
The subtractions are needed because the  \MSbar  scheme 
collinear divergent terms are already included into
the quark  distribution functions $f(x_1,\mu^2)$ and
$f(x_2,\mu^2)$ and into the gluon distribution function $g(x_5,\mu^2)$.
For neutral current processes we have the same formula.

Let us take into consideration the following expression with 
quark-antiquark hard and soft-virtual subtraction terms:
\bqa
&&\int_0^1 d x_1  \int_0^1 d x_2 \, f(x_1,\mu^2) f(x_2,\mu^2)  
\Bigg[ {\hat \sigma}^{\rm CC}_0( x_1 x_2\, s) 
-  {\hat \sigma}^{\rm CC}_{H_{\rm Subtr}}(\mu^2, x_1 x_2\, s)
- {\hat \sigma}^{\rm CC}_{SV_{\rm Subtr}}(\mu^2, x_1 x_2\, s)\Bigg]  
\nll 
&&= \int_0^1 d x_1\,\int_0^1 d x_2\, {\hat \sigma}^{\rm CC}_0( x_1 x_2\, s) \Bigg\{  
f(x_1,\mu^2)\, f(x_2,\mu^2) - \frac{\als}{2 \pi}\,f(x_1,\mu^2) \int_{x_1}^1 d z \, 
f\left(\frac{x_1}{z},\mu^2\right) \nll
&& \times 
\left[ P_{qq}(z) \left(\ln\left(\frac{\mu^2}{m_2^2}\right) 
- 1 - 2 \ln(1-z)\right)\right]_{+}
- \frac{\als}{2 \pi}\,f(x_2,\mu^2) \int_{x_2}^1 d z \,
f\left(\frac{x_2}{z},\mu^2\right) 
\nll && \times
\left[ P_{qq}(z) 
\left(\ln\left(\frac{\mu^2}{m_1^2}\right) - 1 - 2 \ln(1-z)\right)\right]_{+}
       \Bigg\}.
\eqa
We used the ``+'' prescription because we have here difference
${\hat \sigma}^{\rm CC}_0(z,x_1,x_2,s)-{\hat \sigma}^{\rm CC}_0(1,x_1,x_2,s)$.

One can note that we can apply subtractions to the PDF instead of the cross section:
\bqa
&& f(x_i,\mu^2) \to
f(x_i,\mu^2)
- \frac{\als}{2 \pi} \int_{x_i}^1 d z \, 
f\left(\frac{x_i}{z},\mu^2\right)
\nll && \qquad \times
\left[P_{qq}(z)\left(\ln\left(\frac{\mu^2}{m_1^2}\right) 
- 1 - 2 \ln(1-z)\right)\right]_{+}.
\eqa
The same manipulation can be done for the neutral current processes.
In practical applications, if the subtraction is applied to PDFs, one has to
take care on spurious $\order{\alpha_s^2}$ contribution. The can be done by
means of linearization procedure described in Ref.~\cite{Arbuzov:2005dd}.

We have a possibility to integrate numerically  over three  angles 
and over the independent variables $z$, $x_1$ and $x_2$ to obtain 
${\hat \sigma}^{\rm CC}_{\rm Hard}( x_1 x_2\, s)$,  
${\hat \sigma}^{\rm CC}_{q_1g}( x_1 x_5\, s)$ and 
${\hat \sigma}^{\rm CC}_{q_2g}( x_2 x_5\, s)$. 

We introduced a new
variable $W_x = x_1 x_2$ when integrated the hard gluon contribution of
 the quark-antiquark process to obtain  
${\hat \sigma}^{\rm CC}_{\rm Hard}( x_1 x_2\, s)$. If we take $x_2, Wx$ for the 
set of independent variables, the Jacobian of the transition is equal to
$\frac{1}{x_2}$. So we have the following transition of the integrals:
\bqa
\int_0^1 d x_1  \int_0^1 d x_2 \, f(x_1) f(x_2) \qquad  \Longrightarrow
\qquad  \int_0^1 d W_x \int_0^1 d x_2 \,\frac{1}{x_2}\,
 f\left(\frac{W_x}{x_2}\right)\, f(x_2).
\eqa 

When we integrated the hard gluon contribution of
 the quark-gluon process to obtain   
${\hat \sigma}^{\rm CC}_{q_1g}( x_1 x_5\, s) $ or
${\hat \sigma}^{\rm CC}_{q_2g}( x_2 x_5\, s)  $
we introduced a new variable $W_y = x_1 x_5 z$ or $W_y = x_2 x_5 z$. In this 
case we have the following transition of the integrals:
\bqa
\int_0^1 d x_1  \int_0^1 d x_5 \, f(x_1) g(x_5) \int_x^{z_{max}} d z
 \quad  \Longrightarrow
\quad  \int_0^1 d W_y \int_0^1 d x_5  \int_x^{z_{max}} d z \,\frac{1}{x_5 z}\,
 f(\frac{W_y}{x_5 z})\, g(x_5).
\eqa

\section{Conclusions and Numerical Results}

For the sake of comparison with {\tt MCFM}~\cite{Ellis:1999ec}
for numerical evaluations we used the following set of input parameters:
\bq 
\begin{array}[b]{lcllcllcl}
G_F & = & 1.16639 \times 10^{-5} \GeV^{-2}, & \alpha(0) &=& 1/137.03599911, \\
\alpha_s(M_Z) &=& 0.130, & \alpha_s(M_W) &=& 0.1326,  \\
\mw & = & 80.419\GeV, &
\gw & = & 2.06\GeV, \\
\mz & = & 91.188\GeV,& 
\gz & = & 2.49\GeV, \\
\mh & = & 115\GeV, &
m_t & = & 170.9\;\GeV, \\
m_u & = & m_d = 66\;\MeV, &
m_c & = & 1.5\;\GeV, \\
m_s & = & 150\;\MeV, &
m_b & = & 4.62 \;\GeV, \\
|V_{ud}| & = & |V_{cs}| = 0.975, &
|V_{us}| & = & |V_{cd}| = 0.222. 
\end{array}
\label{input}
\eq
The CTEQ6L1~\cite{Pumplin:2002vw} set of PDF~\footnote{The LO PDF were used just for the comparison. For practical applications of the described results NLO PDF should be chosen.} 
was used with the factorization scales
being equal to $\mz$ and $\mw$ for the NC and CC cases, respectively.
The following cuts on the final state kinematics were applied:
\bq
P_t > 25\;\GeV, \qquad
M_{ll} > 20 \;\GeV, \qquad
\eta < 1.2,
\eq
where $P_t$ is the transverse momentum of a lepton,
$M_{ll}$ is the invariant mass of a charged lepton pair (only for NC),
and $\eta$ is the pseudo-rapidity of a charged lepton.

In numerical evaluations we used an adaptive Monte Carlo integrator
based on the VEGAS algorithm~\cite{Lepage:1977sw}. For the partonic
sub-process cross sections we used the standard SANC  FORTRAN modules 
which can either produced interactively by the system or 
just downloaded from the SANC webpage~\cite{SANCwww}. 
These modules are described in Ref.~\cite{Andonov:2008ga}.

\begin{figure}[!ht]
\begin{center}
\includegraphics[width=75mm,height=75mm,angle=90]{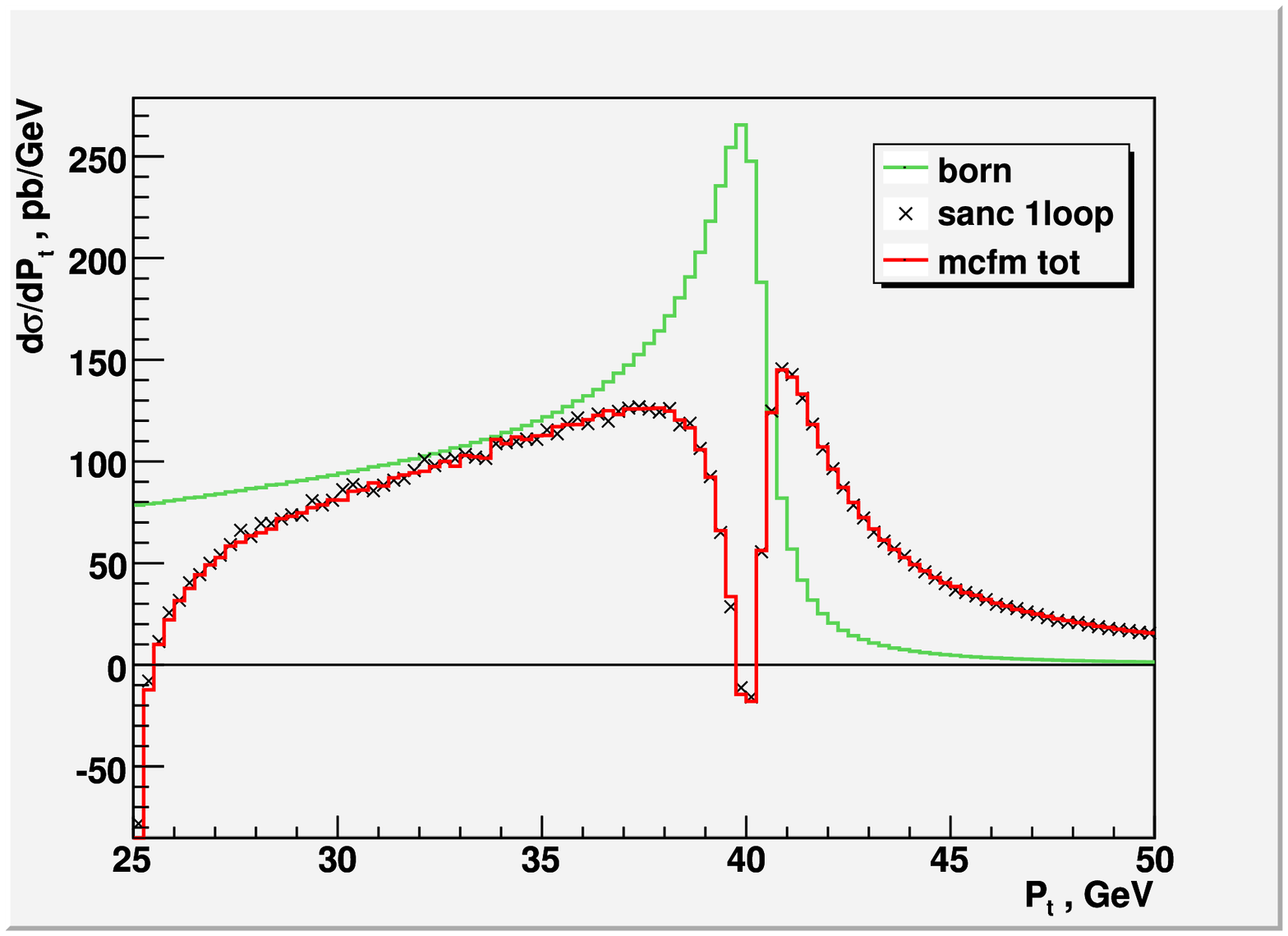} 
\end{center}
\caption{$\mu^+$ transverse momentum distribution in CC Drell--Yan.}
\label{CCpt}
\end{figure}

\begin{figure}[!ht]
\begin{center} \vspace*{1.5cm}
\includegraphics[width=75mm,height=75mm,angle=90]{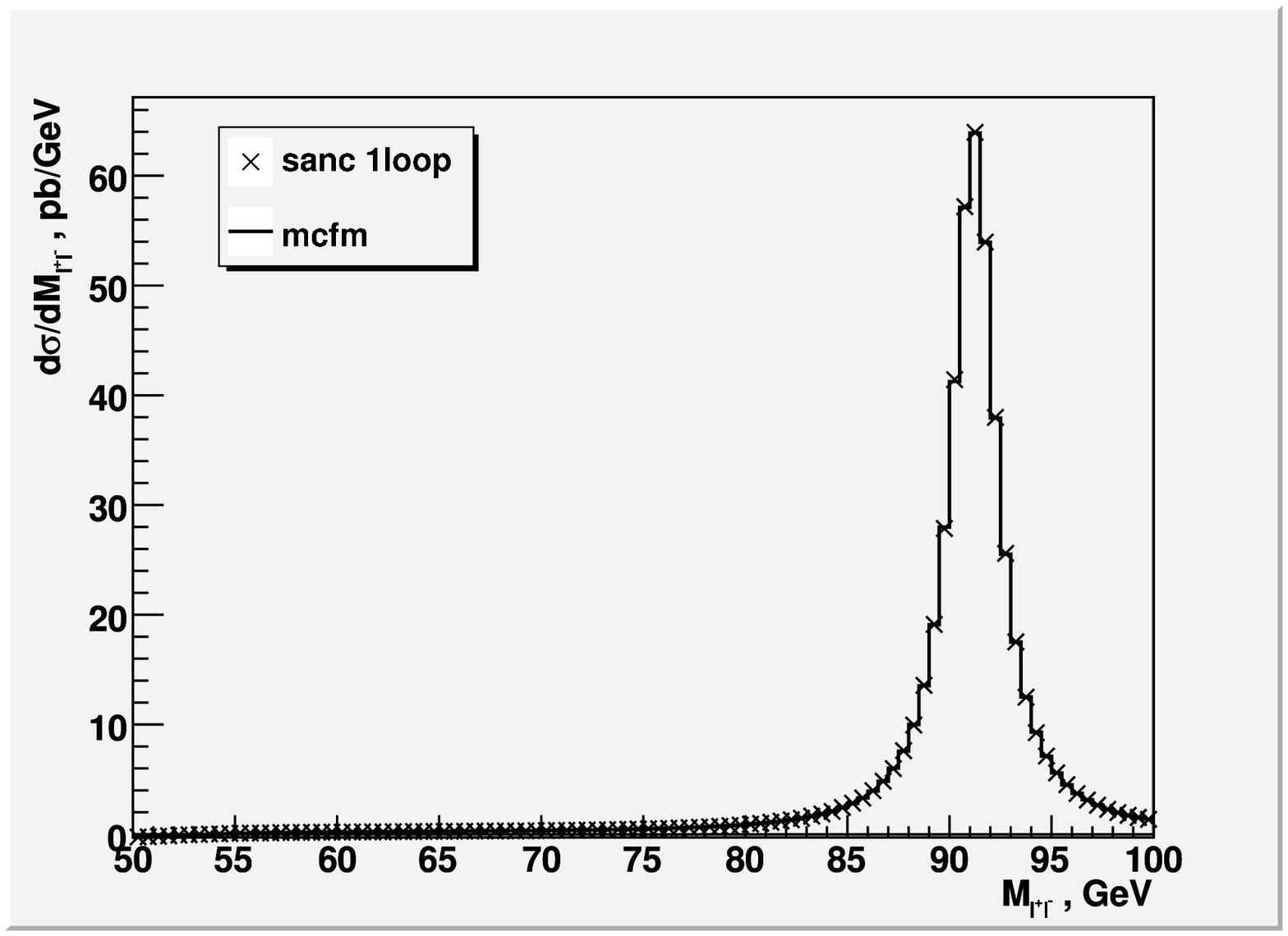} 
\end{center}
\caption{Lepton pair invariant mass distribution in NC Drell--Yan.}
\label{NCminv}
\end{figure}

In Fig.~\ref{CCpt} we show comparison of the SANC results with the MCFM ones 
for the transverse momentum distribution of $\mu^+$ in the charged current
DY process at LHC. Fig.~\ref{NCminv} shows the corresponding comparison of
results for the $\mu^+\mu^-$ invariant mass distribution in the NC case.
Note that the deeps in the first bins of both the distributions are not physical,
they appeared due to kinematical cuts imposed just at the left borders. 
Application of the PDF factorization in SANC and MCFM are performed in different 
schemes: the scheme with massive quarks in SANC (as described above) 
and the \MSbar scheme with massless partons in MCFM. We see that for the given 
distributions the difference between these schemes is not numerically important.

For a realistic application, one has to take into account also QCD showers. 
It can be done with help of the standard packages like PYTHIA~\cite{Sjostrand:2006za} 
and HERWIG~\cite{Corcella:2000bw}. Note that the showers will wash out
the negatively weighted events, which can be seen in the resonance region 
in Fig.~\ref{CCpt}. 

In this way we presented in detail the evaluation of NLO QCD corrections
to Drell--Yan like processes. It is important that we performed it in the
environment of the SANC system, so that now we have a self-consistent 
simultaneous treatment of QCD and electroweak radiative corrections the
the DY processes. It is required for the forthcoming experiments at the LHC.
Simultaneous implementation of the electroweak and NLO QCD corrections to
Drell-Yan processes received with help of the SANC into 
a Monte Carlo event generator will be described elsewhere~\cite{inprep}.

\vspace*{.2cm}

{\bf Acknowledgments} This work was supported by the RFBR grant 07-02-00932.
One of us (A.~Arbuzov) is also grateful to the grant of the RF President 
(Scientific Schools 3312.2008.2).

\end{document}